\documentclass[prb,aps,twocolumn,showpacs]{revtex4}
\usepackage{graphicx}
\usepackage{dcolumn}

\begin{document}

\title{
Vortex states in superconductors with strong Pauli-paramagnetic effect}

\author{Masanori Ichioka}
\affiliation{Department of Physics, Okayama University,
Okayama 700-8530, Japan}
\author{Kazushige Machida}
\affiliation{Department of Physics, Okayama University,
Okayama 700-8530, Japan}
\date{\today}

\begin{abstract}
Using the quasiclassical theory, 
we analyze the vortex structure of strong-paramagnetic superconductors.  
There, induced paramagnetic moments are accumulated 
exclusively around the vortex core.
We quantitatively evaluate the significant paramagnetic effect 
in the $H$-dependence of various quantities, 
such as low temperature specific heat, Knight shift, magnetization 
and the flux line lattice (FLL) form factor. 
The anomalous  $H$-dependence of the FLL form factor 
observed by the small angle neutron scattering in ${\rm CeCoIn_5}$ 
is attributable to the large paramagnetic contribution. 
\end{abstract}


\maketitle
\section{introduction}

Recent two seemingly quite independent activities prompt us 
to cope those in a unified way because two typical experiments 
in each field suggest the vortex structure associated with 
strong influence of the {\it mismatched} Fermi surface, 
namely two Fermi levels for up and down spins are different;
One is in the rotating fermion superfluids 
of neutral $^6$Li atom gases under the population 
imbalance.~\cite{zwierlein, partridge,machida2006,takahashi} 
The other is heavy fermion superconductors 
with Zeeman-shifted Fermi surfaces under an applied field $H$. 
A heavy fermion compound ${\rm CeCoIn_5}$ is 
a prime candidate of a superconductor with strong 
Pauli-paramagnetic effect, 
where at higher fields the upper critical field $H_{\rm c2}$ 
changes to the first order phase transition~\cite{izawa,bianch2002,tayama}  
and new Fulde-Ferrell-Larkin-Ovchinnikov (FFLO) state   
appears.~\cite{bianchi,radovan,watanabe,capan,martin,kakuyanagi,kumagai,matsudaJPSJ}
In the FFLO states, the pair potential is considered to have 
periodic spatial modulation in addition to the vortex 
modulation.\cite{ff,lo,machida1984,tanaka,tachiki,buzdin,adachi2003,ikeda,mizushima,ichiokaFFLO} 
It is because Cooper pairs of up- and down-spins acquire non-zero momentum 
for the center of mass coordinate of the Cooper pair 
by the Fermi surface splitting of up- and down-spin electron bands due 
to Zeeman shift. 
The Fermi surface splitting is also an origin of the Pauli-paramagnetic 
pair breaking and the appearance of paramagnetic moments. 
Therefore, 
even when the vortex states do not enter to the FFLO state yet, 
the strong paramagnetic effects may seriously contribute to 
the vortex state in superconductors.  

  It is expected that, in the presence of strong paramagnetic effect, 
observed quantized vortices in both systems of atomic gases and 
solid states should have universal common 
properties that are absent in conventional vortex picture. 
As for the rotating atomic gas under population imbalance 
within a trap potential, 
the vortex states were studied by Bogoliubov-de Gennes theory 
in the configuration of single vortex.\cite{takahashi}  
There, paramagnetic moments are enhanced in the vortex core region, 
and the vortex core structure is related 
to the spectral evolution of quasiparticles 
around the vortex in the presence of Zeeman shift. 
Thus, also in solid states it is necessary to study the vortex states 
in superconductors with strong paramagnetic effect,  
clarifying the exotic vortex core structure 
of pair potential, paramagnetic moments, internal magnetic 
field distributions, and local electronic states. 
These paramagnetic effects also give significant contributions 
to bulk properties, such as specific heat, 
paramagnetic susceptibility, or  magnetizations.   

In some heavy fermion superconductors, the paramagnetic 
effects due to Zeeman shift are important to understand 
the properties of the vortex states, 
because the superconductivity survives until under high 
magnetic fields due to the effective mass enhancement. 
The paramagnetic contributions are eminent at higher fields. 
For example, the $H$-dependence of low temperature specific heat $C(H)$ 
is often used to distinguish the presence of nodes in 
the pairing potential. 
The Sommerfeld coefficient $\gamma(H)= \lim_{T \rightarrow 0} C(H)/T 
\propto H$ in $s$-wave full gap superconductors, 
and  $\gamma(H) \propto \sqrt{H}$ by the Volovik effect
in $d$-wave pairing with line 
nodes.~\cite{volovik,ichiokaQCLd1,miranovic,nakai}  
The curves of $\gamma(H)$ are expected to smoothly recover 
to the normal state value towards $H_{\rm c2}$. 
However, in some heavy fermion superconductors, 
$C(H)$ deviates from these curves. 
In ${\rm CeCoIn_5}$, $C(H)$ shows concave curves, i.e., 
$C(H) \propto H^\alpha$ ($\alpha >1$) at higher 
fields.\cite{ikedaS,aoki,deguchi}   
This behavior is difficult to be understood only by effects of 
the pairing symmetry.  
A similar $C(H)$ behavior is observed also in 
${\rm UBe_{13}}$.\cite{Ramirez}  
The experimental data of magnetization curve $M_{\rm total}(H)$ 
in ${\rm CeCoIn_5}$ show a concave curve at higher fields, 
instead of a conventional convex curve.\cite{tayama} 
As an anomalous behavior of ${\rm CeCoIn_5}$, 
the small angle neutron scattering (SANS) experiment 
reported the $H$-dependence of flux line lattice (FLL) form factor 
determined from the Bragg intensity.~\cite{DeBeer}  
While the form factor shows exponential decay as a function of $H$ 
in many superconductors, 
it keeps almost constant at lower fields 
for $H \parallel c$ in ${\rm CeCoIn_5}$. 
As for properties of ${\rm CeCoIn_5}$, the contribution of 
antiferromagnetic fluctuation is also proposed 
in addition to the strong paramagnetic 
effect.\cite{matsudaJPSJ,bianchi2003b,paglione,young} 
Therefore, it is expected to study whether 
properties of vortex states in ${\rm CeCoIn_5}$ are 
theoretically explained only by the paramagnetic effect. 
Theoretical studies of the $H$-dependences also  help us to estimate 
strength of the paramagnetic effect 
from experimental data of the $H$-dependences in various superconductors. 

In this paper, 
based on the selfconsistent microscopic calculation of 
quasiclassical Eilenberger 
theory,~\cite{ichiokaQCLd1,miranovic,nakai,eilenberger,kleinJLTP,ichiokaQCLs}
we study the spatial structure of the vortex states 
in the presence of the paramagnetic effect by 
Zeeman-shift.~\cite{ichiokaFFLO,WatanabeKita,klein,adachi}  
We will clarify the paramagnetic effect on the vortex core 
structure, calculating the pair potential, paramagnetic moment, 
internal magnetic field, and local electronic states. 
We also study the paramagnetic effect  
by quantitatively estimating the $H$-dependence 
of low temperature specific heat, 
Knight shift, magnetization and FLL form factors, 
and show the relation of the $H$-dependence behaviors and 
the strength of paramagnetic effect.  
The anomalous field-dependence of FLL form factor 
observed by SANS experiment~\cite{DeBeer} 
is explained by the strong paramagnetic effect. 

Previous works in the selfconsistent quasiclassical calculation were 
mainly applied to the vortex state 
in the absence of paramagnetic effects, and 
successfully estimate local electronic states around the vortex core 
and the $H$-dependence of the low temperature specific heat, 
studying the effect of the pairing 
symmetry,\cite{ichiokaQCLd1,miranovic,ichiokaP} 
gap anisotropy,\cite{nakai} and multi-band structure.~\cite{ichiokaMgB2}   
The selfconsistent calculation of the pair potential is necessary for 
quantitatively valid evaluation of the vortex states, 
since we have to use the correct vortex core size in the calculation 
at each field and temperature. 
As for previous quasiclassical studies on 
the strong paramagnetic effect on the vortex state, 
there were some works focusing on the FFLO vortex 
states.\cite{ichiokaFFLO,klein}  
Without FFLO modulation but with strong paramagnetic effect, 
many studies were done along $H_{\rm c2}(T)$, 
and there were few studies far from $H_{\rm c2}$.  
The quasiclassical study on the $H$-dependence of the specific heat 
and magnetization were done by Adachi {\it et al.} at $T=0.4T_{\rm c}$, 
using Landau level expansion.\cite{adachi}  
In this paper, we study the $H$-dependence at $T=0.1T_{\rm c}$, 
solving the Eilenberger equation numerically by 
the explosion method~\cite{kleinJLTP,ichiokaQCLs}
in the vortex lattice state, and also study the vortex core structure 
including local electronic states, 
and the $H$-dependence of 
paramagnetic susceptibility and the FLL form factors. 

After giving our formulation of quasiclassical theory 
in Sec. \ref{sec:formulation}, 
we study the paramagnetic effect on the $H$-dependence 
of paramagnetic susceptibility and low temperature specific heat 
in Sec. \ref{sec:chi-DOS}. 
In Sec. \ref{sec:vortex-core}, we show the paramagnetic contributions  
on the vortex core structure, and the local electronic state 
in the presence of Zeeman shift.  
In Sec. \ref{sec:FLL}, we estimate the $H$-dependence of FLL form factor, 
and discuss the anomalous $H$-dependence observed by SANS in 
${\rm CeCoIn_5}$. 
The paramagnetic effect on the magnetization curve is presented in 
Sec. \ref{sec:M}, and the last section is devoted to summary and 
discussions. 

\section{Quasiclassical theory including paramagnetic contribution}
\label{sec:formulation}

We calculate the spatial structure of the vortex lattice state 
by quasiclassical Eilenberger theory in the clean 
limit,~\cite{ichiokaQCLd1,miranovic,nakai,eilenberger,kleinJLTP,ichiokaQCLs}
including the paramagnetic effects 
due to the Zeeman term $\mu_{\rm B}B({\bf r})$, 
where $B({\bf r})$ is the flux density of the internal field and 
$\mu_{\rm B}$ is a renormalized Bohr 
magneton.~\cite{ichiokaFFLO,WatanabeKita,klein,adachi} 
The quasiclassical theory is quantitatively valid when 
$\xi \gg 1/k_{\rm F}$ ($k_{\rm F}$ is the Fermi wave number, and 
$\xi$ is the superconducting coherence length), 
which is satisfied in most of superconductors in solid states. 
The quasiclassical Green's functions
$g( \omega_n +{\rm i} {\mu} B, {\bf k},{\bf r})$, 
$f( \omega_n +{\rm i} {\mu} B, {\bf k},{\bf r})$, and 
$f^\dagger( \omega_n +{\rm i} {\mu} B, {\bf k},{\bf r})$  
are calculated in the vortex lattice state  
by the Eilenberger equation 
\begin{eqnarray} &&
\left\{ \omega_n +{\rm i}{\mu}B 
+\tilde{\bf v} \cdot\left(\nabla+{\rm i}{\bf A} \right)\right\} f
=\Delta\phi g, 
\nonumber 
\\ && 
\left\{ \omega_n +{\rm i}{\mu}B 
-\tilde{\bf v} \cdot\left( \nabla-{\rm i}{\bf A} \right)\right\} f^\dagger
=\Delta^\ast \phi^\ast g  , \quad 
\label{eq:Eil}
\end{eqnarray} 
where $g=(1-ff^\dagger)^{1/2}$, ${\rm Re} g > 0$, 
$\tilde{\bf v}={\bf v}/v_{{\rm F}0}$, 
and ${\mu}=\mu_{\rm B} B_0/\pi k_{\rm B}T_{\rm c}$. 
We mainly consider the $d$-wave pairing case   
for a pairing function, $\phi({\bf k})=\sqrt{2}\cos2\theta$, 
as suggested in ${\rm CeCoIn_5}$.\cite{aoki,izawa} 
${\bf k}$ is the relative momentum of the Cooper pair, 
and ${\bf r}$ is the center-of-mass coordinate of the pair. 
We consider the case of two-dimensional 
cylindrical Fermi surface, 
${\bf k}=k_{\rm F}(\cos\theta,\sin\theta)$, 
where $0 \le \theta < 2\pi$. 
The Fermi velocity is given by 
${\bf v}=v_{{\rm F}0}(\cos\theta,\sin\theta)$. 
Throughout this paper, 
length, temperature, and magnetic field are scaled by 
$R_0$, $T_c$, and $B_0$, respectively. 
Here, $R_0=\hbar v_{{\rm F}0}/2 \pi k_{\rm B} T_{\rm c}$, 
$B_0=\hbar c /2|e|R_0^2$. 
Matsubara frequency $\omega_n=(2n+1)\pi T$, energy $E$, 
and pair potential $\Delta$ are scaled by $\pi k_{\rm B} T_{\rm c}$. 
Since magnetic fields are applied to the $z$ axis direction,  
in the symmetric gauge the vector potential 
${\bf A}({\bf r})=\frac{1}{2} \bar{\bf B} \times {\bf r}
 + {\bf a}({\bf r})$, 
where $\bar{\bf B}=(0,0,\bar{B})$ is a uniform flux density and 
${\bf a}({\bf r})$ is related to the internal field 
${\bf B}({\bf r})=\bar{\bf B}+\nabla\times {\bf a}({\bf r})$.
The unit cell of the vortex lattice is given by 
${\bf r}=s_1({\bf u}_1-{\bf u}_2)+s_2{\bf u}_2$ with 
$-0.5 \le s_i \le 0.5$ ($i$=1, 2), ${\bf u}_1=(a,0,0)$ and 
${\bf u}_2=(a/2,a_y,0)$.  
In the $d$-wave pairing, 
we consider the case of square vortex lattice, i.e., 
$a_y/a=1/2$, where the nearest neighbor vortices are located 
to the nodal direction.    
In the $d$-wave superconductors, this square lattice configuration 
is stable at higher fields.~\cite{ichiokaQCLd1,DeBeer}

The pair potential is selfconsistently calculated by 
\begin{eqnarray}
\Delta({\bf r})
= g_0N_0 T \sum_{0 < \omega_n \le \omega_{\rm cut}} 
 \left\langle \phi^\ast({\bf k}) \left( 
    f +{f^\dagger}^\ast \right) \right\rangle_{\bf k} 
\label{eq:scD} 
\end{eqnarray} 
with 
$(g_0N_0)^{-1}=  \ln T +2 T
        \sum_{0 < \omega_n \le \omega_{\rm cut}}\omega_n^{-1} $. 
$\langle \cdots \rangle_{\bf k}$ indicates the Fermi surface average. 
We use $\omega_{\rm cut}=20 k_{\rm B}T_{\rm c}$.
The vector potential for the internal magnetic field 
is selfconsistently determined by 
\begin{eqnarray}
\nabla\times \left( \nabla \times {\bf A} \right) 
=\nabla\times {\bf M}_{\rm para}({\bf r})
-\frac{2T}{{\tilde{\kappa}}^2}  \sum_{0 < \omega_n} 
 \left\langle {\bf v}_{\rm F} 
         {\rm Im} g  
 \right\rangle_{\bf k}, 
\label{eq:scH} 
\end{eqnarray} 
where we consider both the diamagnetic contribution of 
supercurrent in the last term 
and the contribution of the paramagnetic moment 
${\bf M}_{\rm para}({\bf r})=(0,0,M_{\rm para}({\bf r}))$ 
with 
\begin{eqnarray}
M_{\rm para}({\bf r})
=M_0 \left( 
\frac{B({\bf r})}{\bar{B}} 
- \frac{2T}{{\mu} \bar{B} }  
\sum_{0 < \omega_n}  \left\langle {\rm Im} \left\{ g \right\} 
 \right\rangle_{\bf k}
\right) . 
\label{eq:scM} 
\end{eqnarray} 
The normal state paramagnetic moment 
$M_0 = ({{\mu}}/{\tilde{\kappa}})^2 \bar{B} $,   
$\tilde{\kappa}=B_0/\pi k_{\rm B}T_{\rm c}\sqrt{8\pi N_0}$  and 
$N_0$ is the density of states (DOS) 
at the Fermi energy in the normal state. 
We solve Eq. (\ref{eq:Eil}) and Eqs. (\ref{eq:scD})-(\ref{eq:scM})
alternately, and obtain selfconsistent solutions
as in previous works,~\cite{ichiokaQCLs,ichiokaQCLd1,ichiokaMgB2}
under a given unit cell of the vortex lattice. 
The unit cell is divided to $41 \times 41$ mesh-points, 
where we obtain the quasiclassical Green's functions, $\Delta({\bf r})$,
$M_{\rm para}({\bf r})$ and ${\bf A}({\bf r})$. 
When we solve Eq. ({\ref{eq:Eil}) by the explosion
method,
we estimate $\Delta({\bf r})$ and ${\bf A}({\bf r})$ at arbitrary
positions by the interpolation from their values at the mesh points,
and by the periodic boundary condition of the unit cell including the
phase factor due to the magnetic 
field.~\cite{kleinJLTP,ichiokaQCLs,ichiokaQCLd1,ichiokaMgB2}

Using Doria-Gubernatis-Rainer scaling,~\cite{WatanabeKita,doria} 
we obtain the relation of $\bar{B}$ and 
the external field $H$ as 
\begin{eqnarray} && 
H=\left(1-\frac{{\mu}^2}{\tilde{\kappa}^2}\right) 
\left(\bar{B}
+\left\langle \left( B({\bf r})-\bar{B} \right)^2\right\rangle_{\bf r}
/{\bar{B}} \right)
\nonumber \\ &&   
+\frac{T}{\tilde{\kappa}^2 \bar{B}} \langle  \sum_{0 < \omega_n} 
\langle 
{\mu} B({\bf r}) {\rm Im}  \left\{ g \right\}  
+\frac{1}{2}{\rm Re}\left\{ 
\frac{(f^\dagger \Delta+f \Delta^\ast)g}{g+1} \right\} 
\nonumber \\ &&   
\hspace{1cm}
+\omega_l {\rm Re}\{ g-1 \} 
\rangle_{\bf k}\rangle_{\bf r}, 
\end{eqnarray} 
where $\langle \cdots \rangle_{\bf r}$ indicates the spatial average. 
We consider the case of large Ginzburg-Landau (GL) parameter 
$\kappa_{\rm GL} \sim \tilde{\kappa}=89$ 
and low temperature $T/T_{\rm c}=0.1$. 
For the two-dimensional Fermi surface, 
$\tilde{\kappa}
=( 7 \zeta(3) /8 )^{1/2}  \kappa_{\rm GL} 
\sim \kappa_{\rm  GL}$.\cite{miranovic2003}   
In these parameters, $|\bar{B}-H| < 10^{-4} B_0$. 

When we calculate the electronic states,
we solve Eq. (\ref{eq:Eil}) with 
$ {\rm i}\omega_n \rightarrow E + {\rm i} \eta$.
The local density of states (LDOS) is given by
$N({\bf r},E)=N_{\uparrow}({\bf r},E)+N_{\downarrow}({\bf r},E)$, where 
\begin{eqnarray}
N_\sigma({\bf r},E)=N_0 \langle {\rm Re }
\{
g( \omega_n +{\rm i} \sigma{\mu} B, {\bf k},{\bf r})
|_{i\omega_n \rightarrow E + i \eta} \}\rangle_{\bf k}
\end{eqnarray}
with $\sigma=1$ ($-1$) for up (down) spin component. 
We typically use $\eta=0.01$.
The DOS is obtained by the spatial average of the LDOS as 
$N(E)=N_\uparrow (E) +N_\downarrow (E)
 =\langle N({\bf r},E) \rangle_{\bf r}$. 

\section{Field dependence of paramagnetic susceptibility and zero-energy DOS}
\label{sec:chi-DOS}

\begin{figure}[tb]
\includegraphics[width=7.0cm]{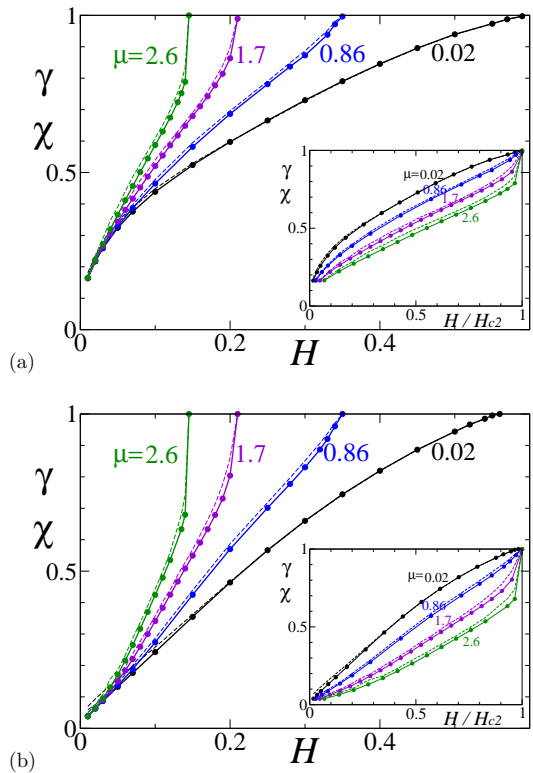}
  \caption{
(color online)
The magnetic field dependence of the paramagnetic susceptibility 
$\chi(H)$ (solid lines) 
and the zero-energy DOS $\gamma(H)$ (dashed lines) at $T=0.1T_{\rm c}$  
for various paramagnetic parameters ${\mu}=0.02$, 0.86, 1.7, and 2.6 
in the $d$-wave (a) and $s$-wave (b) pairing cases. 
The insets show same data as a function of $H/H_{\rm c2}$. 
}
\label{fig:nchi}
\end{figure}

First, we discuss the field dependence of zero-energy DOS 
$\gamma(H)=N(E=0)/N_0$ and paramagnetic susceptibility 
$\chi(H)=\langle M_{\rm para}({\bf r}) \rangle_{\bf r}/M_0$,  
which are normalized by the normal state values. 
From low temperature specific heats $C$, we obtain 
$\gamma(H) \propto C/T$ experimentally. 
And $\chi(H)$ is observed by the Knight shift in NMR experiments,  
which measure the paramagnetic component induced by an external field 
via the hyperfine coupling between a nuclear spin and conduction electrons. 
In $d$-wave pairing, 
$\chi(H)$ shows $\sqrt{H}$-behavior at low fields.~\cite{zheng} 

As shown in Fig. \ref{fig:nchi}, $\gamma$ (dashed lines) and  
$\chi$ (solid lines) show almost the same behavior 
at low temperatures. 
First, we see the case of $d$-wave pairing with line nodes in 
Fig.  \ref{fig:nchi}(a). 
There $\gamma(H)$ and $\chi(H)$ describe $\sqrt{H}$-like recovery smoothly 
to the normal state value ($\gamma=\chi=1$ at $H_{\rm c2}$)
in the case of negligible paramagnetic effect 
(${\mu}=0.02$).~\cite{volovik,ichiokaQCLd1,miranovic,nakai}  
With increasing the paramagnetic parameter ${\mu}$, 
$H_{\rm c2}$ is suppressed and 
the Volovik curve  $\gamma(H)\propto\sqrt{H}$ gradually changes 
into curves with a concave curvature. 
For large ${\mu}$, $H_{\rm c2}$ changes to first order 
phase transition.~\cite{adachi} 
We note that at lower fields all curves 
exhibit a $\sqrt{H}$ behavior because the paramagnetic effect 
($\propto H$) is not effective. 
Further increasing $H$, $\gamma(H)$ behaves quite differently. 
There we find a turning point field 
which separates a convex curve at lower $H$ 
and a concave curve at higher $H$. 
This inflection point increases as ${\mu}$ decreases.
In the inset of Fig. \ref{fig:nchi}, we plot $\gamma(H)$ 
and $\chi(H)$ as a function of $H/H_{\rm c2}$. 
The overall $H$-dependence at $0<H< H_{\rm c2}$ 
can be used to analyze experimental data,  
in order to estimate the strength of the paramagnetic effect, 
${\mu}$. 

To examine effects of the pairing symmetry, 
we show  $\gamma(H)$ and $\chi(H)$ also for $s$-wave pairing 
$\phi({\bf k})=1$ in Fig.  \ref{fig:nchi}(b), 
where we use the triangular lattice configuration $a_y/a=\sqrt{3}/2$. 
In the $H$-dependence of  $\gamma(H)$ and $\chi(H)$, 
differences by the vortex lattice configuration of 
triangular or square are small. 
The difference in the $H$-dependences of
Figs. \ref{fig:nchi}(a) and \ref{fig:nchi}(b) at low fields 
comes from the gap structure of the pairing function. 
In the full gap case of $s$-wave pairing, 
$\gamma(H)$ and $\chi(H)$ show $H$-linear-like behavior 
at low fields.  
With increasing the paramagnetic effect, 
$H$-linear behaviors gradually change  
into curves with a concave curvature. 
As seen in  Figs. \ref{fig:nchi}(a) and \ref{fig:nchi}(b), 
paramagnetic effects appear similarly at high fields  
both for $s$-wave and $d$-wave pairings.  

The concave curves of the specific heat 
at higher fields by strong paramagnetic effect were also 
presented in Ref. \onlinecite{adachi} at $T=0.4T_{\rm c}$. 
In the present calculation, the concave curves are confirmed 
even at low temperature $T=0.1T_{\rm c}$, where $C/T \sim \gamma(H)$ 
without large specific heat jump at $H_{\rm c2}$. 
In $\gamma(H)$ at low $T$, the differences of $s$-wave and $d$-wave 
pairings at lower fields are clearly seen.

\section{Paramagnetic contribution on vortex core structure}
\label{sec:vortex-core}

\begin{figure}[tb]
\includegraphics[width=8.5cm]{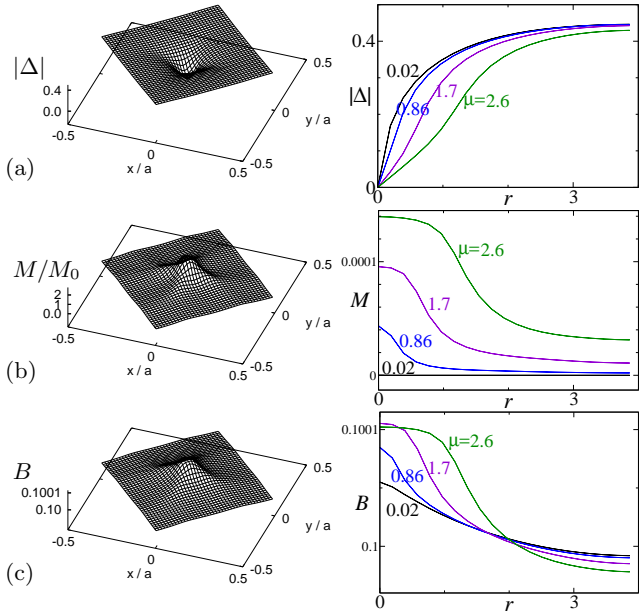}
  \caption{
(color online)
Spatial structure of the pair potential (a), 
paramagnetic moment (b) and internal magnetic field (c) 
at $T=0.1T_{\rm c}$ and $\bar{B}=0.1B_0$, where $a=11.2R_0$,  
in $d$-wave pairing. 
The left panels show $|\Delta({\bf r})|$, $M_{\rm para}({\bf r})$, 
and $B({\bf r})$ within a unit cell of the square vortex lattice 
at ${\mu}=1.7$. 
The right panels show the profiles along the trajectory $r$ from the 
vortex center to the midpoints between nearest neighbor vortices. 
${\mu}=0.02$, 0.86, 1.7, and 2.6.  
}
 \label{fig:prf}
\end{figure}

In order to understand the contribution of the paramagnetic effect 
on the vortex structure, 
we illustrate the local structures of 
the pair potential $|\Delta({\bf r})|$, 
paramagnetic moment $M_{\rm para}({\bf r})$, and 
internal magnetic field $B({\bf r})$ 
within a unit cell of the vortex lattice in Fig. \ref{fig:prf}.  
Since we assume $d$-wave pairing with the line node gap here, 
the vortex core structure is deformed to fourfold symmetric shape 
around a vortex core.~\cite{ichiokaQCLd1,ichioka1996} 
It is noted that the paramagnetic moment is enhanced exclusively around 
the vortex core, as shown in Fig. \ref{fig:prf}(b) where the four ridges of
paramagnetic moment  are extended towards 
the anti-nodal directions from the core. 
Since the contribution of the paramagnetic vortex core is enhanced 
with increasing ${\mu}$,  
internal field $B({\bf r})$ consisting of diamagnetic 
and paramagnetic contributions  
is further enhanced around the vortex core by the paramagnetic effect, 
as shown in Fig. \ref{fig:prf}(c).  
When ${\mu}$ is large, 
the pair potential $|\Delta({\bf r})|$ is slightly suppressed around 
the paramagnetic vortex core, and the vortex core radius is enlarged,   
as shown in Fig. \ref{fig:prf}(a).

\begin{figure}[tb]
\includegraphics[width=8.5cm]{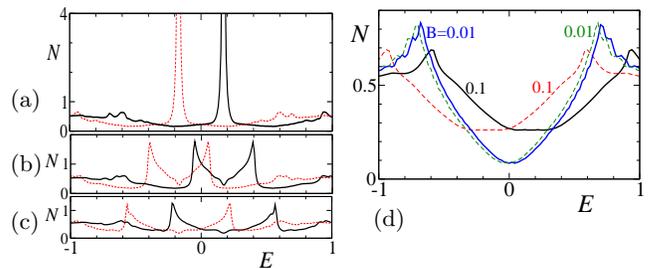}
  \caption{
(color online)
Local density of states at $r/R_0=0$ (a), $0.8$ (b) 
and $1.6$ (c)  
from the vortex center towards the nearest neighbor vortex direction 
in $d$-wave pairing.  
Solid lines show $N_\uparrow ({\bf r},E)/N_0$ for up-spin electrons,  
and dashed lines show $N_\downarrow ({\bf r},E)/N_0$ at $\bar{B}=0.1B_0$.  
$\mu=1.7$ and $T=0.1T_{\rm c}$. 
(d) Spatial-averaged DOS at $\bar{B}/B_0=0.1$ and $0.01$
in $d$-wave pairing.  
Solid lines show $N_\uparrow (E)/N_0$ for up-spin electrons,  
and dashed lines show $N_\downarrow (E)/N_0$. 
}
 \label{fig:ldos}
\end{figure}

The enhancement of $M_{\rm para}({\bf r})$ around vortex core 
is related to spatial structure of the LDOS 
$N_\sigma ({\bf r},E)$. 
As shown in Fig. \ref{fig:ldos}(a), 
the LDOS spectrum shows {\it zero-energy peak} at the vortex center, 
but the spectrum is shifted to $E=\pm {\mu} \bar{B}$ 
due to Zeeman shift. 
The peak states at $E > 0$ is empty for $N_\uparrow(E,{\bf r})$, 
and the peak at $E < 0$ is occupied  for $N_\downarrow(E,{\bf r})$.  
Therefore, from the relation 
\begin{eqnarray}
M_{\rm para}({\bf r})=-\mu_{\rm B}\int_{-\infty}^0
(N_\uparrow(E,{\bf r}) -N_\downarrow(E,{\bf r}) ){\rm d}E, 
\label{eq:M-LDOS}
\end{eqnarray}
large $M_{\rm para}({\bf r})$ appears due to the local 
imbalance of up- and down-spin occupation around the vortex core. 
As shown in Figs. \ref{fig:ldos}(b) and \ref{fig:ldos}(c), 
increasing the distance from the vortex center, 
the peak of the spectrum is split into two peaks, 
which are shifted to higher and lower energies, respectively. 
When one of split peaks crosses $E=0$, 
the imbalance of up- and down-spin occupation is decreased. 
Thus, $M_{\rm para}({\bf r})$ is suppressed outside of vortex cores. 

In Fig. \ref{fig:ldos}(d), 
we present the spectrum of spatially-averaged DOS. 
In the DOS spectrum, 
peaks of the LDOS are smeared by the spatial average.  
Because of the flat spectrum at low energies,   
paramagnetic susceptibility $\chi(H)$ shows 
almost the same $H$-behavior as the zero-energy DOS 
$\gamma(H) \sim N(E=0)$ even for large ${\mu}$, 
as shown in Fig. \ref{fig:nchi}, 
while $\chi(H)$ counts the DOS contribution 
in the energy range $|E|<{\mu}H$, i.e., 
$\chi(H)\sim \int_0^{\mu H} N_\uparrow(E){\rm d}E/\mu H$ 
from Eq. (\ref{eq:M-LDOS}). 

\section{Field dependence of flux line lattice form factor}
\label{sec:FLL}

\begin{figure}[tb]
\includegraphics[width=7.0cm]{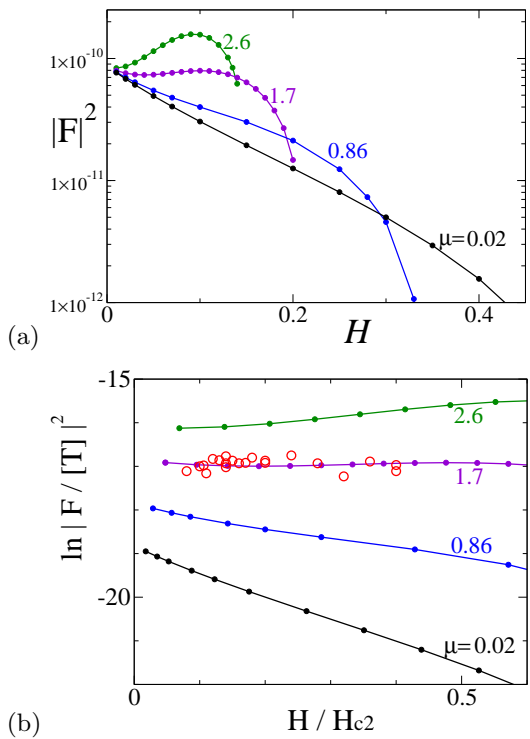}
  \caption{
(color online)
Field dependence of FLL form factor $F_{1,0}$ 
for ${\mu}=0.02$, 0.86, 1.7, and 2.6 at $T=0.1T_{\rm c}$ 
in $d$-wave pairing.   
(a) $|F_{1,0}|^2$ is plotted as a function of $H$.  
The vertical axis is in logarithmic scale. 
(b) We plot $\ln |F_{1,0}|^2$ as a function of $H/H_{\rm c2}$, 
where $F_{1,0}$ is scaled so that $H_{\rm c2}$ for each $\mu$ 
corresponds to 5 [T]. 
Open circles are experimental data in ${\rm CeCoIn_5}$ 
observed by SANS experiment.\cite{DeBeer} 
}
 \label{fig:ff}
\end{figure}

One of the best ways to directly see the accumulation of 
the paramagnetic moment around the vortex core 
is to observe the Bragg scattering intensity 
of the FLL via SANS experiment.  
The intensity of the $(h,k)$-diffraction peak is given by 
$ I_{h,k}=|F_{h,k}|^2/|{\bf q}_{h,k}| $ 
with the wave vector ${\bf q}_{h,k}=h{\bf q}_1+k{\bf q}_2$, 
${\bf q}_1=(2 \pi/a, -\pi/a_y,0)$ and   
${\bf q}_2=(2 \pi/a, \pi/a_y,0)$. 
The Fourier component $F_{h,k}$ is given by 
$B({\bf r})=\sum_{h,k}F_{h,k}
\exp({\rm i}{\bf q}_{h,k}\cdot{\bf r})$. 
The intensity of the main peak at $(h,k)=(1,0)$ in the 
SANS for FLL observation probes 
the magnetic field contrast between the vortex cores and the surrounding.

We calculate the field dependence of $|F_{1,0}|^2$, 
which is shown in Fig. \ref{fig:ff}. 
In the case of negligible paramagnetic effect (${\mu}=0.02$), 
$|F_{1,0}|^2$ decreases exponentially as a function of $H$. 
This is a result for the $d$-wave pairing at low $T$ 
in the clean limit. 
With increasing paramagnetic effect,  
the decreasing slope of $|F_{1,0}|^2$ becomes gradual, 
and changes to increasing functions of $H$ at lower fields  
in extremely strong paramagnetic case (${\mu}=2.6$).    
This is because $|F_{1,0}|$ includes enhanced paramagnetic 
contribution proportional to ${\mu}H$, reflecting the 
enhanced internal field around the vortex core, 
shown in Fig. \ref{fig:prf}(c), by the paramagnetic moment. 

The SANS experiment in ${\rm CeCoIn_5}$ reported that 
$|F_{1,0}|^2$ is almost constant as a function of $H$ within 
the field range $0.08 \le H/H_{c2} \le 0.4$ 
(0.4[T] $\le  H \le$ 2.0[T]) for $H \parallel c$.~\cite{DeBeer} 
This behavior is reproduced by our calculation for ${\mu}\sim1.7$. 
There, $|F_{1,0}|^2$ shows flat behavior at low fields, since 
the paramagnetic contribution increasing with $H$ compensates 
the conventional decrease of $|F_{1,0}|^2$ as a function of $H$. 
For the comparison to the experimental data, 
we plot $\ln |F_{1,0}|^2$ as a function of $H/H_{\rm c2}$ at 
lower fields in Fig. \ref{fig:ff}(b). 
There, unit of the magnetic field in the calculated data for each $\mu$  
is rescaled so that $H_{\rm c2}$ corresponds to 5 [T], i.e., 
$H_{\rm c2}$  in ${\rm CeCoIn_5}$ for $H \parallel c$. 
For quantitative accordance of the results for $\mu=1.7$ with   
the experimental data, 
we tune the GL parameter as $\tilde{\kappa}=89$. 
The variations of internal fields are roughly 
proportional to $\tilde{\kappa}^{-2}$, 
as seen from Eq. (\ref{eq:scH}).   
Changing $\tilde{\kappa}$, we can shift curves in Fig. \ref{fig:ff}(b) 
towards the vertical direction.  
The slopes of the curves in Fig. \ref{fig:ff}(b) are determined by 
the paramagnetic effect. 
With increasing $\mu$, the negative slope becomes gradual, and 
even changes to positive slope.  
When the paramagnetic effect is negligible ($\mu=0.02$), 
$\ln |F_{1,0}|^2$ decreases by 2 in the field 
range $0.1 < H/H_{\rm c2}< 0.5 $, which corresponds to 
the exponential decay as a function of $H$, 
as expected in conventional superconductors 
(also see the theoretical curves in Ref. \onlinecite{DeBeer}). 
For the large paramagnetic case $\mu=1.7$,  
$\ln |F_{1,0}|^2$ does not decrease as a function of $H$, 
which accords with the experimental data 
[circles in  Fig. \ref{fig:ff}(b)] of SANS experiments.  
The anomalous $H$-dependence of the SANS intensity in 
${\rm CeCoIn_5}$ can be explained by the strong paramagnetic effect, 
and suggests   ${\mu}\sim 1.7$.   
When ${\mu}\sim 1.7$, $H_{\rm c2}$ is about $38 \% $ suppressed by 
the paramagnetic pair breaking from the value of 
no paramagnetic effect, as seen in Fig. \ref{fig:nchi}.   
The strong paramagnetic contributions are also considered as 
an origin of new FFLO phase and 
first order $H_{\rm c2}$ phase transition at higher fields.  

The purpose of this discussion was to demonstrate that 
the paramagnetic effect can change the slope of 
$\ln |F_{1,0}|^2$ as a function of $H$, and that 
in the case of strong paramagnetic effect 
$|F_{1,0}|^2$ does not show exponential decay.  
We note that for further nice fitting to the experimental data, 
there is a room to include the effect by the Fermi surface anisotropy 
or by deformations of the vortex lattice configuration.

\section{Field dependence of Magnetization}
\label{sec:M}

\begin{figure}[tb]
\includegraphics[width=7.0cm]{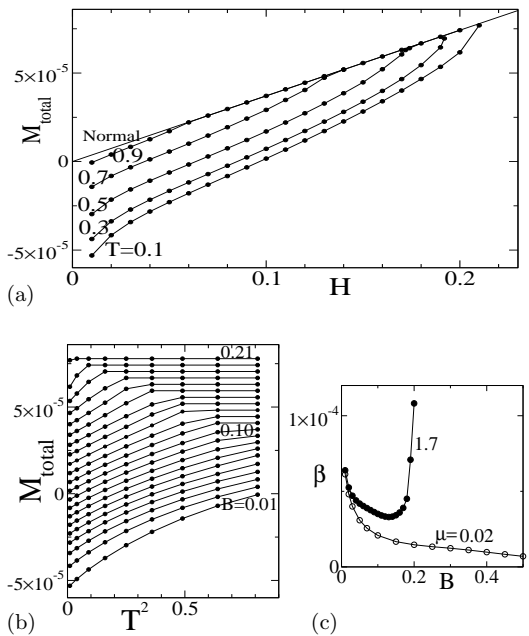}
  \caption{
(color online)
(a) 
Magnetic field dependence of magnetization 
$M_{\rm total}$ for ${\mu}=1.7$ at 
$T/T_{\rm c}=0.1$, 0.3, 0.5, 0.7, 0.9 and 1.0 (normal state) 
in $d$-wave pairing. 
(b) 
$M_{\rm total}$ as a function of $T^2$ at ${\bar B}=0.01$, 0.02, 0.03, 
$\cdots$, 0.21.
(c) $H$-dependence of factor $\beta(H)$ at ${\mu}=0.02$ and 1.7. 
}
 \label{fig:mtotal}
\end{figure}

We discuss the paramagnetic effect on the magnetization curves.
In Fig. \ref{fig:mtotal}(a), 
magnetization curves are presented as a function of $H$ 
for various $T$ at ${\mu}=1.7$. 
The magnetization $M_{\rm total}=\bar{B}-H$ includes both 
the diamagnetic and the paramagnetic contributions. 
It is seen that $M_{\rm total}(H)$ exhibits a sharp rise 
near $H_{\rm c2}$ by the paramagnetic pair breaking effect, 
and that $M_{\rm total}(H)$ has concave curvature at higher fields, 
instead of a conventional convex curvature. 
These behaviors are seen in experimental data of 
${\rm CeCoIn_5}$,~\cite{tayama} 
and previous calculation at $T=0.4T_{\rm c}$.\cite{adachi}  

In Fig. \ref{fig:mtotal}(b), $M_{\rm total}$ is plotted as a function 
of $T^2$ for various $\bar{B}$. 
We fit these curves as 
$M_{\rm total}(T,H)=M_0+\frac{1}{2}\beta(H)T^2+O(T^3)$ at low $T$. 
The slope $\beta(H)
=\lim_{T \rightarrow 0} \partial^2 M_{\rm total}/\partial T^2 $ 
decreases on raising $H$ at lower fields. 
However, at higher fields approaching $H_{\rm c2}$, 
the slope $\beta(H)$ sharply increases. 
Thus, as shown in Fig. \ref{fig:mtotal}(c),  
$\beta(H)$ as a function of $H$ exhibits a minimum  at 
intermediate $H$ and rapid increase near $H_{\rm c2}$ 
by the paramagnetic effect when ${\mu}=1.7$. 
This is contrasted with the case of negligible 
paramagnetic effect (${\mu}=0.02$), 
where $\beta(H)$ is a decreasing function of $H$ until $H_{\rm c2}$.

The behavior of $\beta(H)$ is consistent with that of $\gamma(H)$, 
since there is a relation $\beta(H)\propto \partial \gamma(H)/\partial H$ 
obtained from a thermodynamic Maxwell's relation 
$\partial^2 M_{\rm total}/\partial T^2 = \partial (C/T)/\partial B$ 
and $B \sim H$.~\cite{adachi} 
In Fig. \ref{fig:nchi}, we see that  for ${\mu}=1.7$ 
the slope of $\gamma(H)$ is decreasing function of $H$ at low $H$, 
but changes to increasing function near $H_{\rm c2}$. 
This behavior correctly reflects the $H$-dependence of $\beta(H)$. 
The rapid increase near $H_{\rm c2}$ is clearly seen at lower temperatures, 
compared with the results at higher temperatures.\cite{adachi} 

\section{Summary and discussions}

We studied the vortex states in the presence of strong paramagnetic effect 
by selfconsistent quasiclassical calculations, 
which can be used for quantitative estimate of the vortex states 
even far from $H_{\rm c2}$. 
Calculating the spatial structure of the vortex states and 
local electronic states, we clarified the paramagnetic effects 
on the vortex core structure. 
There, the core radius is enlarged and 
the internal field around the core is further enhanced, 
due to the enhanced paramagnetic moments at the vortex core.

Qualitatively estimating the $H$-dependence of 
low temperature specific heat, Knight shift, magnetization, 
and FLL form factor, 
we showed the relationship between the $H$-dependence behaviors and 
the strength of the paramagnetic effect.  
The specific heat, Knight shift, and magnetization show 
rapid increase near $H_{\rm c2}$, 
due to the paramagnetic pair breaking 
which is eminent at higher fields. 
The anomalous $H$-dependences of FLL form factor of SANS 
experiment~\cite{DeBeer} in ${\rm CeCoIn_5}$ are also explained by  
the strong paramagnetic effect. 
This reflects the paramagnetic vortex core structure, 
affecting the internal field distribution. 
These theoretical studies of the $H$-dependences 
help us to evaluate the strength of the paramagnetic effect 
from the experimental data of the $H$-dependences 
in various superconductors.  
For example, analyses for ${\rm Sr_2RuO_4}$, ${\rm TmNi_2B_2C}$ and 
${\rm URu_2 Si_2}$ are given 
elsewhere.\cite{machidaSRO,DeBeerTmNiBC,YanoURuSi}

As for ${\rm CeCoIn_5}$, 
our analysis of FLL form factor in SANS experiment suggests 
${\mu}\sim 1.7$.  
This indicates strong paramagnetic effect, 
so that $H_{\rm c2}$ is about $38 \% $ suppressed by 
the paramagnetic pair breaking from the value of no paramagnetic effect.  
On the other hand, when we compare the $H$-dependence of the 
specific heat with the experimental data,\cite{deguchi} 
experimental data $\gamma(H) \sim  C/T$ also show the concave curve 
at higher fields, as suggested in our calculation. 
However, at lower fields, 
$\gamma(H)$ is much smaller than that expected by 
our theoretical calculation.   
This discrepancy indicates that 
experimental data may include other additional contributions, 
such as antiferromagnetic fluctuations,  or 
$H$-dependent bulk properties other than the conduction electrons. 
Therefore, in order to understand the  $H$-dependences 
of the vortex state properties in ${\rm CeCoIn_5}$, 
we need further careful studies by the 
collaboration of experimental and theoretical studies.

\section*{Acknowledgments} 
 
The authors are grateful for useful discussions and communications with 
T. Mizushima, H. Adachi, N. Nakai, K. Kumagai, 
and M.R. Eskildsen.


\end{document}